\documentclass[conference, a4paper]{IEEEtran}
\IEEEoverridecommandlockouts
\usepackage[sort]{cite}
\usepackage{amsmath,amssymb,amsfonts}
\usepackage{algorithmic}
\usepackage{graphicx}
\usepackage{hyperref}
\hypersetup{
    colorlinks=true,
    citecolor=blue,
    linkcolor=black,
    linkbordercolor={1 1 1},
    citebordercolor={1 1 1},
}
\usepackage{textcomp}
\usepackage{url}
\usepackage{xcolor}
\def\BibTeX{{\rm B\kern-.05em{\sc i\kern-.025em b}\kern-.08em
    T\kern-.1667em\lower.7ex\hbox{E}\kern-.125emX}}
    
\usepackage{balance}

\usepackage[pscoord]{eso-pic}
\newcommand{\placetextbox}[3]{
\setbox0=\hbox{#3}
\AddToShipoutPictureFG*{
\put(\LenToUnit{#1\paperwidth},\LenToUnit{#2\paperheight}){\vtop{{\null}\makebox[0pt][c]{#3}}}%
}%
}%

\usepackage{tikz}
\usetikzlibrary{positioning}

\begin{document}

\title{Dialogue Enhancement and Listening Effort in Broadcast Audio: A Multimodal Evaluation}

\author{\IEEEauthorblockN{Matteo Torcoli, Thomas Robotham, Emanu{\"e}l A.~P.~Habets}

\IEEEauthorblockA{
\textit{International Audio Laboratories Erlangen, Germany*\thanks{*A joint institution of the Friedrich-Alexander-Univerist{\"a}t Erlangen-N{\"u}rnberg (FAU) and Fraunhofer IIS, Erlangen, Germany.}}\\
\{matteo.torcoli, thomas.robotham, emanuel.habets\}@audiolabs-erlangen.de
}}

\maketitle
\begin{abstract}
Dialogue enhancement (DE) plays a vital role in broadcasting, enabling the personalization of the relative level between foreground speech and background music and effects. DE has been shown to improve the quality of experience, intelligibility, and self-reported listening effort (LE). A physiological indicator of LE known from audiology studies is pupil size. The relation between pupil size and LE is typically studied using artificial sentences and background noises not encountered in broadcast content.
This work evaluates the effect of DE on LE in a multimodal manner that includes pupil size (tracked by a VR headset) and real-world audio excerpts from TV. Under ideal listening conditions, 28 normal-hearing participants listened to 30 audio excerpts presented in random order and processed by conditions varying the relative level between foreground and background audio. One of these conditions employed a recently proposed source separation system to attenuate the background given the original mixture as the sole input. After listening to each excerpt, subjects were asked to repeat the heard sentence and self-report the LE. Mean pupil dilation and peak pupil dilation were analyzed and compared with the self-report and the word recall rate. The multimodal evaluation shows a consistent trend of decreasing LE along with decreasing background level.
DE, also when enabled by source separation, significantly reduces the pupil size as well as the self-reported LE.
This highlights the benefit of personalization functionalities at the user’s end.

\end{abstract}
\placetextbox{0.5}{0.08}{\fbox{\parbox{\dimexpr\textwidth-2\fboxsep-2\fboxrule\relax}{\footnotesize \centering Accepted paper. \copyright  2022 IEEE. Personal use of this material is permitted. Permission from IEEE must be obtained for all other uses, in any current or future media, including reprinting/republishing this material for advertising or promotional purposes, creating new collective works, for resale or redistribution to servers or lists, or reuse of any copyrighted component of this work in other works.}}}

\begin{tikzpicture}[overlay, remember picture]
\path (current page.north) node (anchor) {};
\node [below=of anchor] {%
Accepted to 2022 14th International Conference on Quality of Multimedia Experience (QoMEX)};
\end{tikzpicture}

\section{Introduction}

Difficulties in following speech due to loud background sounds have been a known issue in television for over 20 years~\cite{Mathers:1991, shirley:2004, fuchs:2012, uhle:2008, IBC:2021}. The optimal relative level between speech and background is very personal~\cite{torcoli2019preferred, geary2020loudness} and significant listening effort (LE) can be experienced even if intelligibility is perfect~\cite{klink2012measuring, rennies2014listening}. Dialogue enhancement (DE) allows the user to adjust the relative speech level to suit individual needs and preferences. DE can be provided by Next Generation Audio (NGA), e.g., MPEG-H Audio~\cite{Simon:2019}, and has been shown to improve the intelligibility~\cite{fuchs:2014} and the quality of experience, also when enabled by source separation~\cite{torcoli2018adjustment, IBC:2021}. However, these studies did not analyze LE. 

Although LE may be understood at an intuitive level, no unanimous consensus has been reached within audiology disciplines regarding its conceptual scope~\cite{pichora2016hearing}. It remains unclear whether LE is a single concept or if it is an umbrella term for multiple phenomena. It is hypothesized that different measures of LE potentially tap into different underlying dimensions, highlighting the importance of a multimodal evaluation~\cite{alhanbali2019measures}. The main methods to measure LE are:
\begin{enumerate}
\item Self-report, i.e., rating the perceived effort on a scale.
\item Behavioral, i.e., performance on one or more given tasks.
\item Physiological, e.g., electroencephalography, skin conductance, and pupil size.
\end{enumerate}

Pupil size was shown to have an inverse relation with the signal-to-noise ratio (SNR) while listening to speech over noise~\cite{kramer1997assessing, klink2012measuring, zekveld2018pupil, woodcock2019pupil}.
This is usually studied under conditions not encountered in TV, e.g., starting from very low intelligibility levels or using artificial types of noise (white, bubble, speech-shaped) and sentences (e.g., the Harvard Sentences\footnote{\url{https://www.cs.columbia.edu/~hgs/audio/harvard.html}}). It is unclear if pupillometry can also be used under test conditions that allow for increased ecological validity, e.g., using real-world TV material.
A few recent works investigated LE considering material and conditions closer to the TV application~\cite{huber2020asr,westhausen2021reduction,mcclenaghan2022next}, but they did not consider pupillometry. 

This work presents a multimodal evaluation of LE on real-world TV audio excerpts. Self-report, word recall rate (as the behavioral measure), and pupil size are recorded and analyzed.
An off-the-shelf VR headset is used for pupillometry: besides offering full control on the luminance \cite{eckert2021cognitive}, this opens up future research directions, e.g., including video and VR applications.

\begin{figure}[t]
\centering
        \includegraphics[width=0.85\columnwidth]{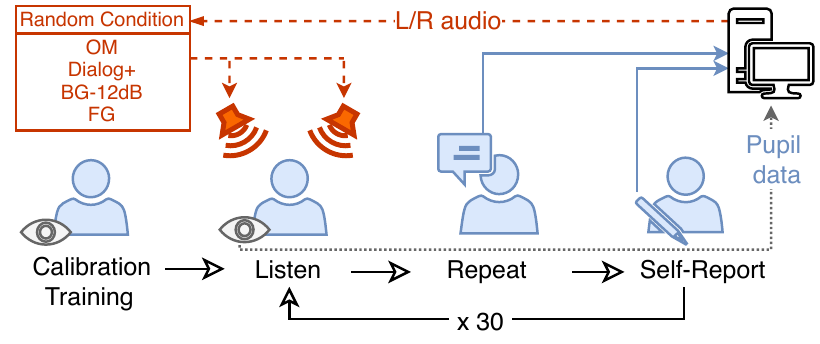}%
        \vspace*{-3mm}
        \caption{
        Test methodology: after an initial phase for calibration and training, the participants were asked to listen one time to each of the 30 audio excerpts, repeat the heard sentence, and self-report the listening effort. While listening, the pupil size was tracked by an off-the-shelf VR headset. Each audio excerpt presented was processed by a random condition varying the relative level between foreground speech and background noise, music, and effects.
        \vspace*{-3mm}
        }
\label{fig:procedure}
\end{figure}

\section{Test Methodology}

As depicted in Figure\,\ref{fig:procedure}, the adopted test procedure consisted of three main phases: listening, repeating, and self-report. Each participant in the test listened to 30 audio excerpts with no repetitions. After each excerpt, they were asked to repeat all the words they could memorize and report how difficult it was to understand the heard sentence.
Before starting, the set-up and the procedure were illustrated, and the participants provided consent for collecting the required data. Two extra audio excerpts were used during the training phase (not considered in the following results), and it was explained how the VR headset (HTC Vive Pro Eye) works and how it can be adjusted to be as comfortable as possible. The eye calibration built in the VR headset was performed with each participant. The full experiment session (including calibration and training) took between 20 and 30 minutes.

The participants in the evaluation were 28 colleagues and students. They were German native speakers, aged between 21 and 46 years (median age was 30; 10 participants were female), with no known hearing impairments. The expertise level was heterogeneous. Almost all work on audio-related topics and had previously participated in listening tests. Only a small minority had previous experience with VR headsets. 

Each participant was asked to sit on a non-revolving chair without head support. The chair was positioned in front of two studio monitors, approximately at the height of the listener's head. The participant and the loudspeakers formed an equilateral triangle with sides of 2\,m. The room was quiet and partially acoustically treated. The VR environment consisted of a monochromatic space, including only a virtual interface used to rate the LE. For this, the 5-point P.800 scale~\cite{P800} translated to German was used. In-between values could also be selected, making the scale continuous in practice.

As audio stimuli, 30 stereo excerpts from German TV were selected (sampling frequency 48\,kHz), for which the original audio stems were available. The accompanying video was not used. The excerpts from TV documentaries and movies had a length of 8 seconds, and a speech-to-background ratio of [-6, 6] dB. The number of words spoken in the excerpts was in the range [7, 29], without counting articles. The excerpts included varying types of voices, paces, speech clarity, emotions, types of background music, and effects. With this heterogeneous real-world audio material, we take a step towards a more ecologically valid test. The video should be considered in the future to take a further step in this direction.

For each participant, the excerpts were presented in random order. For each participant and excerpt, a random processing condition was selected, i.e., each participant was presented with only one excerpt/condition combination and listened to each sentence only once. This is fundamental for evaluating LE, as listening to the same sentence a second time would make it easier to understand, regardless of the processing condition. There were four processing conditions:

\textbf{1) OM}: The unprocessed stereo original mixture from TV.

\textbf{2) Dialog+}: This condition takes the OM and inputs it to the source separation system in~\cite{IBC:2021, strauss:21,torcoli2021controlling,paulus22sampling}.
This is a fully convolutional deep neural network, trained on 48\,kHz high-quality TV material. The estimated background component is remixed with a $-12$\,dB factor together with the estimated foreground component. This simulates the case in which the original audio stems are not available from production.

\textbf{3) BG-12dB}: This condition uses the original foreground speech (FG) and background (BG) stems, simulating a native NGA production.
FG and BG are mixed with a $-12$\,dB factor applied to the BG. (In OM, they are mixed with a 1:1 ratio.)

\textbf{4) FG}: FG only is presented and BG is discarded entirely.

All audio excerpts were loudness normalized to -23\,LUFS before and after processing, and the playback level was kept constant over the full session and for all participants. Pink noise at -23\,LUFS resulted in 59\,dBA at the listening position.
Our main interest was comparing OM and Dialog+. These two conditions were twice as likely to be selected during the condition randomization. This means that the final sample size for OM and Dialog+ is twice as big as BG-12dB and FG.

\section{Multimodal Evaluation Data Analysis}

The self-report was simply stored without post-processing. As a behavioral measure, the sentence repeated by the participant was compared against the ground-truth transcript and a word recall rate (WRR) was computed. Articles were ignored as well as the exact order of the words.

The pupil size while listening was recorded using the eye-tracking system built in the VR headset. This provided a stream of observations over time that was analyzed as follows. The mean over both eyes was considered. A delay of 0.5\,s was applied to compensate for the pupil reaction time \cite{eckert2021cognitive, ellis1981pupillary}. 
To discard unrealistically fast changes in pupil size (due to measurement noise), we removed the samples for which we observed changes larger than 2 times the standard deviation (computed over the full session) with respect to the previous sample, similarly to~\cite{zhang2021disentangling}.
Following~\cite{woodcock2019pupil}, a first-order interpolation was applied to bridge missing points resulting from the previous data cleaning, blinking, and occlusions.

The pupil baseline was computed as the mean over a 0.5\,s window before audio playback. Mean pupil dilation (MPD) and peak pupil dilation (PPD) were computed with respect to the baseline. Both MPD and PPD were transformed into \mbox{z-scores} to account for individual variability~\cite{eckert2021cognitive}, obtaining z\_MPD and z\_PPD.
The relative luminance of the head-mounted display was also tracked. This was $0.13$ throughout the full session, with negligible variations.

\section{Results}

\begin{figure*}[ht]
\centering
        \includegraphics[width=0.2496\linewidth]{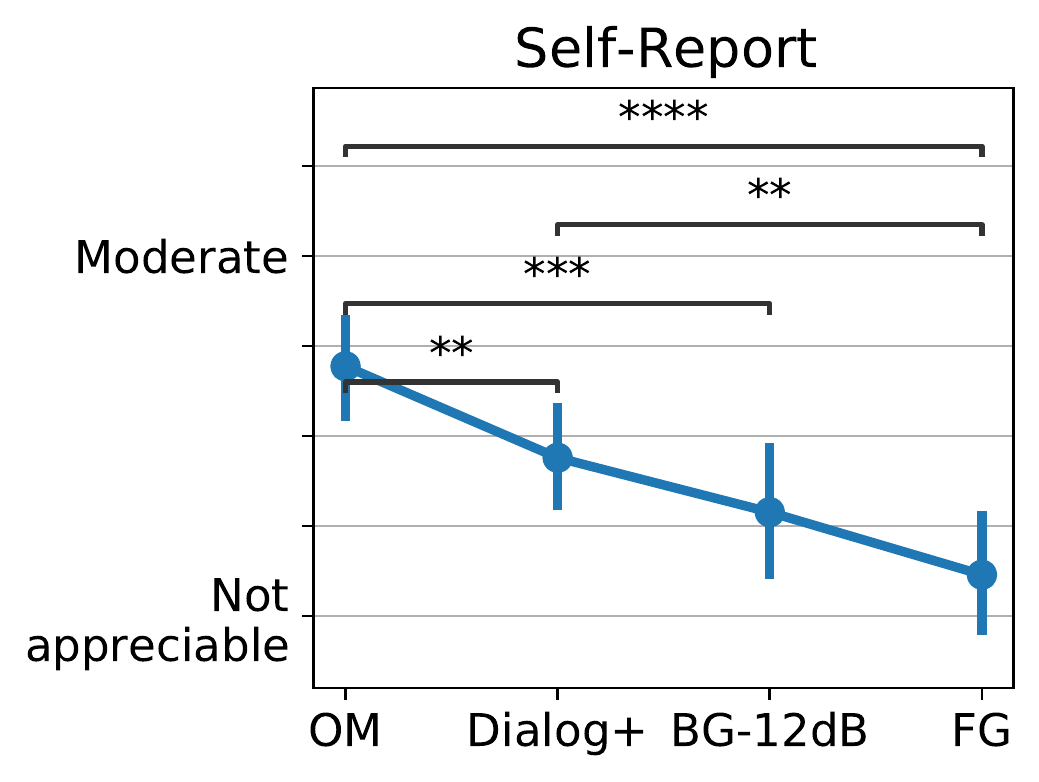}%
        \hfill
        \includegraphics[width=0.2490\linewidth]{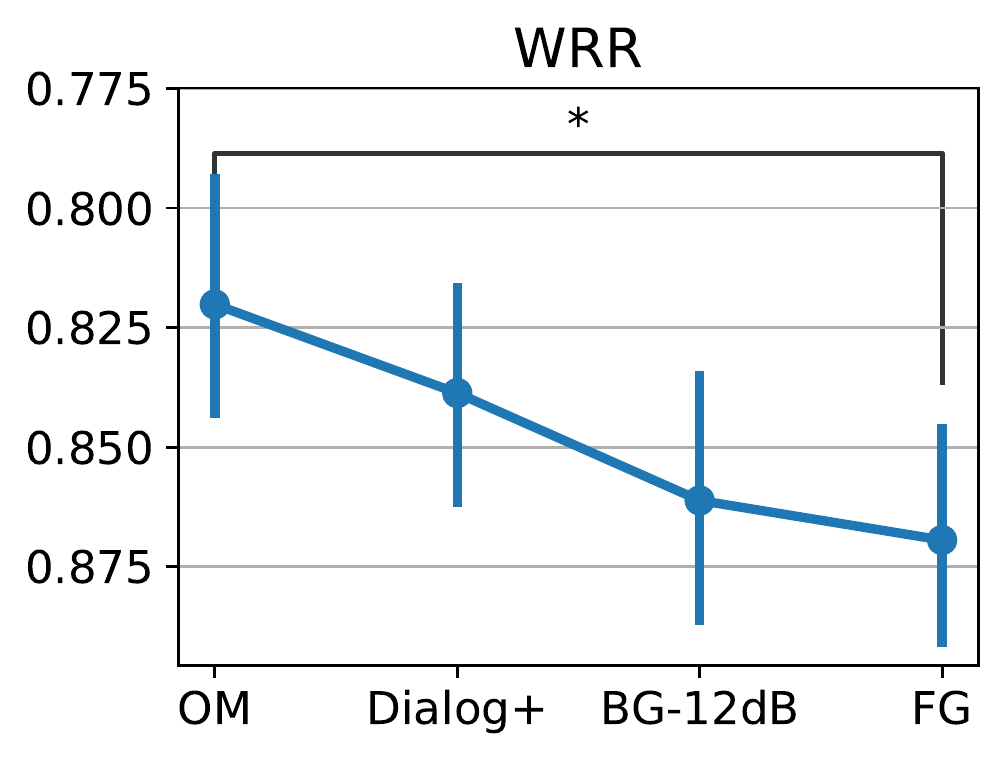}%
        \hfill
        \includegraphics[width=0.248\linewidth]{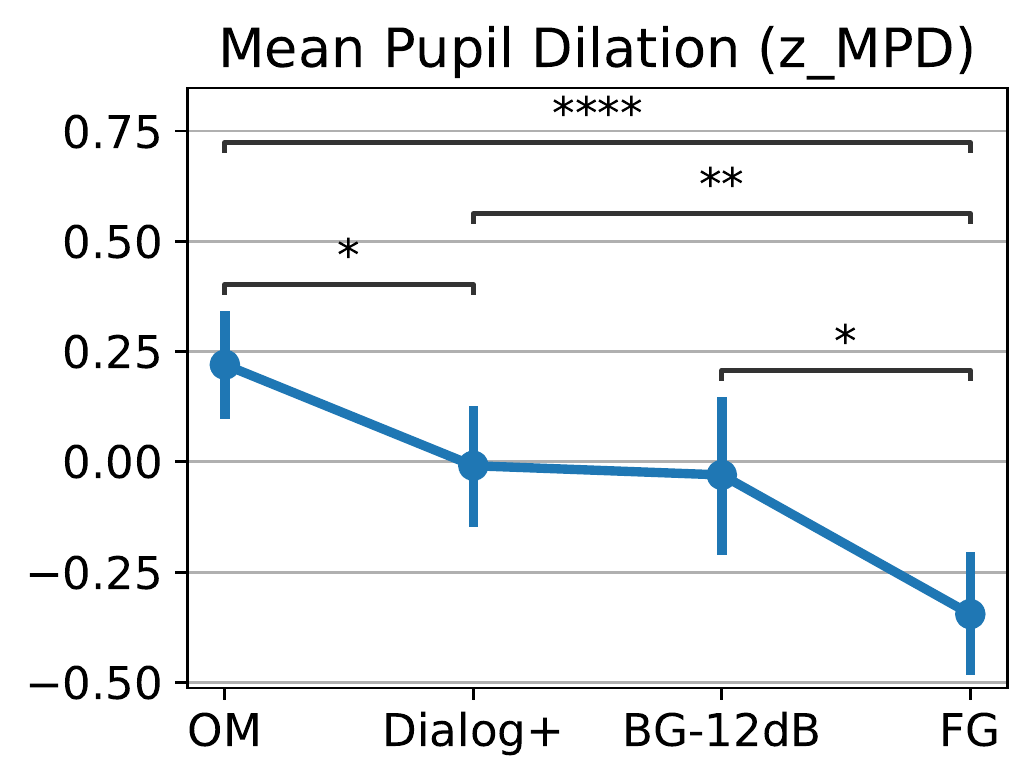}%
        \hfill
        \includegraphics[width=0.241\linewidth]{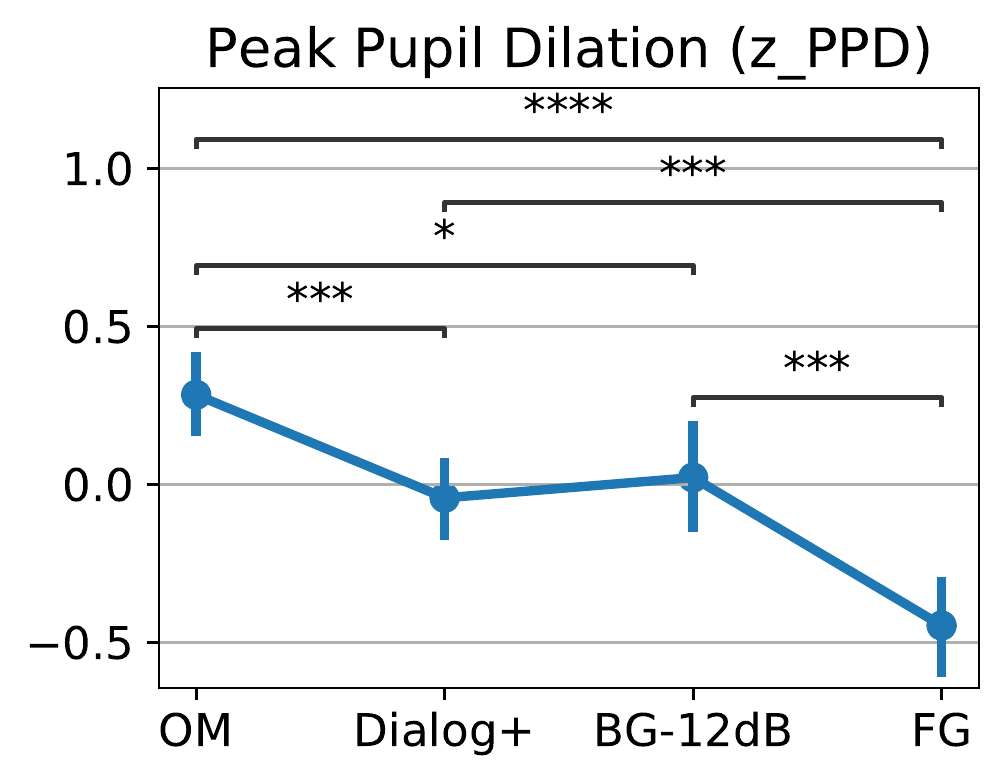}%
        \hfill
        \vspace*{-3mm}
        \caption{
        Main results depicted as mean over the 28 participants and 95\% confidence intervals. Significantly different pairs are shown with asterisk notation, i.e.: *: $0.05 > p > 0.01$. **: $0.01 >= p > 0.001$. ***: $0.001 >= p > 0.0001$. ****: $p <= 0.0001$.
        For better visibility, only 2 points of the 5-point effort scale are shown, and the labels are abbreviated. 
        The original labels were: 1) \textit{Complete relaxation possible; no effort required.} 2) \textit{Attention necessary; no appreciable effort required.} 3) \textit{Moderate effort required.} 4) \textit{Considerable effort required.} 5) \textit{No meaning understood with any feasible effort.
        \vspace*{-3mm}
        }
        }
\label{fig:main}
\end{figure*}

Figure\,\ref{fig:main} shows the main results. An analysis was performed using different statistical tools\footnote{
For self-report, conditions are not normally distributed (Shapiro's test), so Kruskal-Wallis H-test is selected (independent variable: \textit{condition}, dependent variable: \textit{self-report}, $p<10^{-6}$), and followed by Dunn's post-hoc. 
For WRR, conditions are normally distributed, but not homoscedastic (Levene's test). Welch's ANOVA ($p=0.02$) and Games-Howells post-hoc are performed.

For z\_MPD and z\_PPD, normality and homoscedasticity of the conditions are verified. One-way ANOVAs are performed for independent variable \textit{condition} on dependent variable \textit{z\_MPD} and \textit{z\_PPD}, returning $p<10^{-6}$ for both, and followed by a Tukey's HSD tests (95\% confidence).
}. 
Pairs that were found to be significantly different are indicated in Figure\,\ref{fig:main}.

All measures show a clear trend: the highest LE is observed for OM. This is then reduced by lower relative background levels (Dialog+ and BG-12dB), followed by the least LE measured for speech only (FG).
The difference between OM and FG is statistically significant for all measures, suggesting that background music and effects introduce significant LE in TV audio.
DE reduces LE, both via the original stems (\mbox{BG-12dB}) and via blind source separation (Dialog+). In particular, Dialog+ shows a significant difference with OM in terms of z\_MPD, z\_PPD, and self-report.
This is the first time that source separation is shown to reduce pupil size on real-world TV audio.
\hbox{BG-12dB} behaves similarly to Dialog+ (no statistically significant difference observed between them).

The similar trends observed across self-report, WRR, and pupil size measures support that the multimodal approach taken in this study can be employed in an experimental set-up representative of real-world settings and stimuli, using off-the-shelf hardware, while still obtaining consistent results overall.

WRR shows the biggest variance and suffers the most from real-world stimuli and sentences. In many cases, participants could guess or reconstruct the real-world sentences based on semantics and independently of the level of the background, which is undesired in WRR studies. Additionally, some errors were not considered, e.g., a common mistake was to repeat \textit{"He was \textbf{not} married"} while the correct transcript was \textit{"He was married"}. As all words in the ground-truth transcript are repeated, both correspond to 100\% WRR.

\begin{figure}[t]
\centering
        \includegraphics[width=0.9\columnwidth]{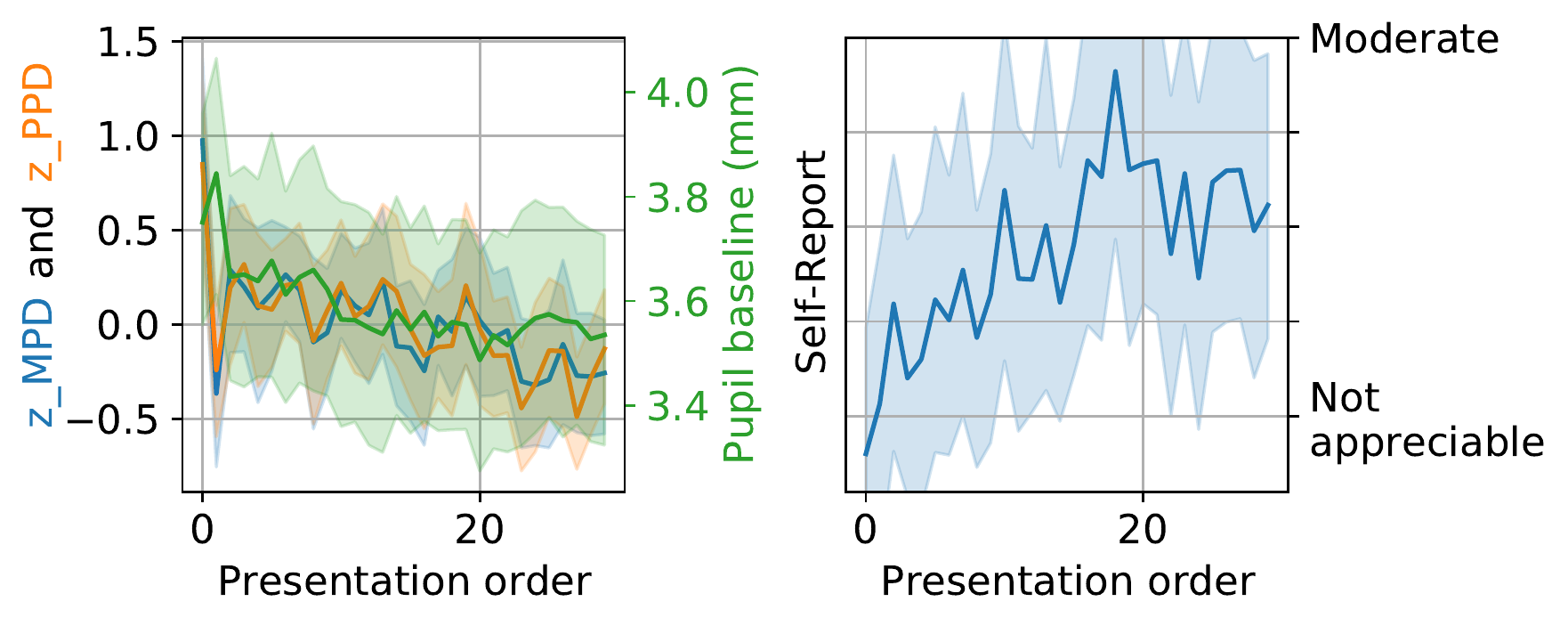}%
        \vspace*{-3mm}
        \caption{
        Effect of presentation order: mean and 95\% confidence intervals over all participants and audio excerpts presented as $i$-th, with $i \in [0, 1, ... 29]$ on the \mbox{x-axis}. The decreasing pupil size and increasing self-report could be explained by an initial familiarization with the task at hand, followed by gradually increasing fatigue and possibly decreasing engagement.
        \vspace*{-3mm}
        }
\label{fig:trend}
\end{figure}

A clear effect of the presentation order is observed on the pupil size and the self-report (Figure\,\ref{fig:trend}), while no effect is observed in terms of WRR (hence not shown in the figure).
Considering the full session, the pupil size measures exhibit a steep decrease at the beginning (even though this is preceded by a training phase) followed by a steady decline.
The decrease at the beginning is also observed in literature and explained as possibly due to participants getting comfortable with tasks and devices~\cite{zekveld2018pupil, borghini2018listening,zekveld2010pupil}.
The following steady decline would suggest a slow decrease in mental workload.
Self-report shows an opposite (increasing) trend, suggesting increasing fatigue.
In \cite{wang2018relations}, it is argued that a more fatigued individual will expend less, not more, resources to achieve the same task. Fatigued individuals may be less motivated to perform well in the test and will exert less effort to perform the task, resulting in a reduction in pupil size, and an increased self-report.

\section{Conclusions and Future Works}

In TV, background music and effects introduce significant LE, even under ideal listening conditions. Decreasing the background level via DE, also via source separation, reduces the LE, as evaluated in a multimodal manner, including self-report, word recall rate, and pupil size.
Background music and effects can carry vital information~\cite{ward2017effect} and play an important role in engaging and entertaining the audience. Yet, they come with a LE cost, requiring care at the production stage (see, e.g., the recommendations in~\cite{geary2020loudness}), and personalization functionalities at the user's end~\cite{Simon:2019,IBC:2021, mcclenaghan2022next}.

The similar overall trends observed across measures indicate that a multimodal evaluation that includes pupillometry can be employed with stimuli representative of real-life TV content. Finer scale effects, e.g., the effect of presentation order, can be better understood when looking at different measures accommodating the multidimensional nature of LE.
The effect of presentation order calls for particular caution while designing experiments of this type. The test phase order (e.g., first repeat, then self-report) might also have an entanglement effect, as well as the natural memory skills of the participants and their engagement with the content. These are points that should be studied in the future. We considered 8\,seconds long excerpts, while pupillometry offers the possibility to analyze longer periods, e.g., the full duration of a movie. This is left for future works, for which this study lays the foundation.

\section*{Acknowledgment}

The authors would like to thank the numerous colleagues who participated in the test and gave valuable feedback, with special thanks to Marie Eckert and Olli S. Rummukainen.

\balance
\bibliographystyle{IEEEtran}
\bibliography{IEEEabrv,references}

\begin{thebibliography}{10}
\providecommand{\url}[1]{#1}
\csname url@samestyle\endcsname
\providecommand{\newblock}{\relax}
\providecommand{\bibinfo}[2]{#2}
\providecommand{\BIBentrySTDinterwordspacing}{\spaceskip=0pt\relax}
\providecommand{\BIBentryALTinterwordstretchfactor}{4}
\providecommand{\BIBentryALTinterwordspacing}{\spaceskip=\fontdimen2\font plus
\BIBentryALTinterwordstretchfactor\fontdimen3\font minus
  \fontdimen4\font\relax}
\providecommand{\BIBforeignlanguage}[2]{{%
\expandafter\ifx\csname l@#1\endcsname\relax
\typeout{** WARNING: IEEEtran.bst: No hyphenation pattern has been}%
\typeout{** loaded for the language `#1'. Using the pattern for}%
\typeout{** the default language instead.}%
\else
\language=\csname l@#1\endcsname
\fi
#2}}
\providecommand{\BIBdecl}{\relax}
\BIBdecl

\bibitem{Mathers:1991}
C.~D. Mathers, ``{A Study of Sound Balances for the Hard of Hearing},''
  \emph{BBC R\&D White Paper, Report 1991-03}, 1991.

\bibitem{shirley:2004}
B.~G. Shirley and P.~Kendrick, ``{ITC Clean Audio Project},'' in \emph{116th
  Audio Engineering Society Convention}, Berlin, Germany, 2004.

\bibitem{fuchs:2012}
H.~Fuchs, S.~Tuff, and C.~Bustad, ``{Dialogue Enhancement - Technology and
  Experiments},'' EBU Technical Review - Q2, Tech. Rep., 2012.

\bibitem{uhle:2008}
C.~Uhle, O.~Hellmuth, and J.~Weigel, ``{Speech Enhancement of Movie Sound},''
  in \emph{125th Audio Engineering Society Convention}, San Francisco, USA,
  2008.

\bibitem{IBC:2021}
M.~Torcoli, C.~Simon, J.~Paulus, D.~Straninger, A.~Riedel, V.~Koch, S.~Wirts,
  D.~Rieger, H.~Fuchs, C.~Uhle, S.~Meltzer, and A.~Murtaza, ``{Dialog+ in
  Broadcasting: First Field Tests using Deep-Learning Based Dialogue
  Enhancement},'' in \emph{Int.~Broadcast.~Conv.~(IBC) Technical Papers}, 2021.

\bibitem{torcoli2019preferred}
M.~Torcoli, A.~Freke-Morin, J.~Paulus, C.~Simon, and B.~Shirley, ``{Preferred
  Levels for Background Ducking to Produce Esthetically Pleasing Audio for TV
  with Clear Speech},'' \emph{Journal of the Audio Engineering Society},
  vol.~67, no.~12, pp. 1003--1011, 2019,
  \url{https://doi.org/10.17743/jaes.2019.0052}.

\bibitem{geary2020loudness}
D.~Geary, M.~Torcoli, J.~Paulus, C.~Simon, D.~Straninger, A.~Travaglini, and
  B.~Shirley, ``{Loudness Differences for Voice-over-Voice Audio in TV and
  Streaming},'' \emph{Journal of the Audio Engineering Society}, vol.~68,
  no.~11, pp. 810--818, 2020, \url{https://doi.org/10.17743/jaes.2020.0022}.

\bibitem{klink2012measuring}
K.~B. Klink, M.~Schulte, and M.~Meis, ``{Measuring Listening Effort in the
  Field of Audiology — A Literature Review of Methods (part 1)},''
  \emph{Z.~Audiol.}, vol.~51, no.~2, pp. 60--67, 2012.

\bibitem{rennies2014listening}
J.~Rennies, H.~Schepker, I.~Holube, and B.~Kollmeier, ``{Listening Effort and
  Speech Intelligibility in Listening Situations Affected by Noise and
  Reverberation},'' \emph{The Journal of the Acoustical Society of America},
  vol. 136, no.~5, pp. 2642--2653, 2014,
  \url{https://doi.org/10.1121/1.4897398}.

\bibitem{Simon:2019}
C.~Simon, M.~Torcoli, and J.~Paulus, ``{MPEG-H Audio for Improving
  Accessibility in Broadcasting and Streaming},'' \emph{arXiv:1909.11549},
  2019.

\bibitem{fuchs:2014}
H.~Fuchs and D.~Oetting, ``{Advanced Clean Audio Solution: Dialogue
  Enhancement},'' \emph{SMPTE Motion Imaging Journal}, vol. 123, no.~5, pp.
  23--27, 2014, \url{https://doi.org/10.5594/j18429}.

\bibitem{torcoli2018adjustment}
M.~Torcoli, J.~Herre, H.~Fuchs, J.~Paulus, and C.~Uhle, ``{The
  Adjustment/Satisfaction Test (A/ST) for the Evaluation of Personalization in
  Broadcast Services and its Application to Dialogue Enhancement},''
  \emph{{IEEE} Transactions on Broadcasting}, vol.~64, no.~2, pp. 524--538,
  2018, \url{https://doi.org/10.1109/TBC.2018.2832458}.

\bibitem{pichora2016hearing}
M.~K. Pichora-Fuller, S.~E. Kramer, M.~A. Eckert, B.~Edwards, B.~W. Hornsby,
  L.~E. Humes, U.~Lemke, T.~Lunner, M.~Matthen, C.~L. Mackersie \emph{et~al.},
  ``{Hearing Impairment and Cognitive Energy: The Framework for Understanding
  Effortful Listening (FUEL)},'' \emph{Ear and Hearing}, vol.~37, pp. 5S--27S,
  2016, \url{https://doi.org/10.1097/AUD.0000000000000312}.

\bibitem{alhanbali2019measures}
S.~Alhanbali, P.~Dawes, R.~E. Millman, and K.~J. Munro, ``{Measures of
  Listening Effort are Multidimensional},'' \emph{Ear and Hearing}, vol.~40,
  no.~5, pp. 1084--1097, 2019,
  \url{https://doi.org/10.1097/AUD.0000000000000697}.

\bibitem{kramer1997assessing}
S.~E. Kramer, T.~S. Kapteyn, J.~M. Festen, and D.~J. Kuik, ``{Assessing Aspects
  of Auditory Handicap by Means of Pupil Dilatation},'' \emph{Audiology},
  vol.~36, no.~3, pp. 155--164, 1997,
  \url{https://doi.org/10.3109/00206099709071969}.

\bibitem{zekveld2018pupil}
A.~A. Zekveld, T.~Koelewijn, and S.~E. Kramer, ``{The Pupil Dilation Response
  to Auditory Stimuli: Current State of Knowledge},'' \emph{Trends in hearing},
  vol.~22, pp. 1--25, 2018, \url{https://doi.org/10.1177/2331216518777174}.

\bibitem{woodcock2019pupil}
J.~S. Woodcock, B.~M. Fazenda, T.~J. Cox, W.~J. Davies \emph{et~al.}, ``{Pupil
  Dilation Reveals Changes in Listening Effort due to Energetic and
  Informational Masking},'' in \emph{23rd International Congress on Acoustics
  (ICA)}, Aachen , Germany, 2019, pp. 6193--6198.

\bibitem{huber2020asr}
R.~Huber, H.~Baumgartner, S.~Goetze, and J.~Rennies, ``{ASR-Based, Single-Ended
  Modeling of Listening Effort - A Tool for TV Sound Engineers},'' in
  \emph{Forum Acusticum}, 2020, pp. 2441--2445.

\bibitem{westhausen2021reduction}
N.~L. Westhausen, R.~Huber, H.~Baumgartner, R.~Sinha, J.~Rennies, and B.~T.
  Meyer, ``{Reduction of Subjective Listening Effort for TV Broadcast Signals
  With Recurrent Neural Networks},'' \emph{{IEEE/ACM} Transactions on Audio,
  Speech, and Language Processing}, vol.~29, pp. 3541--3550, 2021,
  \url{https://doi.org/10.1109/TASLP.2021.3126931}.

\bibitem{mcclenaghan2022next}
I.~McClenaghan, L.~Pardoe, and L.~Ward, ``{The Next Generation of Audio
  Accessibility},'' in \emph{152nd Audio Engineering Society Convention}, The
  Hague, Netherlands, 2022.

\bibitem{eckert2021cognitive}
M.~Eckert, E.~A. Habets, and O.~S. Rummukainen, ``{Cognitive Load Estimation
  Based on Pupillometry in Virtual Reality with Uncontrolled Scene Lighting},''
  in \emph{13th IEEE International Conference on Quality of Multimedia
  Experience (QoMEX)}, Montreal, Canada, 2021, pp. 73--76,
  \url{https://doi.org/10.1109/QoMEX51781.2021.9465417}.

\bibitem{P800}
{International Telecommunication Union}, ``{Recommendation ITU--T P.800 Methods
  for Subjective Determination of Transmission Quality},'' 1996.

\bibitem{strauss:21}
M.~Strauss, J.~Paulus, M.~Torcoli, and B.~Edler, ``{A Hands-On Comparison of
  DNNs for Dialog Separation Using Transfer Learning from Music Source
  Separation},'' in \emph{INTERSPEECH}, Brno, Czech Republic, 2021, pp.
  3900--3904, \url{https://doi.org/10.21437/Interspeech.2021-1418}.

\bibitem{torcoli2021controlling}
M.~Torcoli, J.~Paulus, T.~Kastner, and C.~Uhle, ``{Controlling the Remixing of
  Separated Dialogue with a Non-Intrusive Quality Estimate},'' in \emph{{IEEE}
  Workshop on Applications of Signal Processing to Audio and Acoustics
  (WASPAA)}, New Paltz, NY, USA, 2021, pp. 91--95,
  \url{https://doi.org/10.1109/WASPAA52581.2021.9632756}.

\bibitem{paulus22sampling}
J.~Paulus and M.~Torcoli, ``{Sampling Frequency Independent Dialogue
  Separation},'' in \emph{30th European Signal Processing Conference
  (EUSIPCO)}, Belgrade, Serbia, 2022.

\bibitem{ellis1981pupillary}
C.~Ellis, ``{The Pupillary Light Reflex in Normal Subjects.}'' \emph{British
  Journal of Ophthalmology}, vol.~65, no.~11, pp. 754--759, 1981,
  \url{https://doi.org/10.1136/bjo.65.11.754}.

\bibitem{zhang2021disentangling}
Y.~Zhang, A.~Lehmann, and M.~Deroche, ``{Disentangling Listening Effort and
  Memory Load Beyond Behavioural Evidence: Pupillary Response to Listening
  Effort During a Concurrent Memory Task},'' \emph{PloS one}, vol.~16, no.~3,
  pp. 1--24, 2021, \url{https://doi.org/10.1371/journal.pone.0233251}.

\bibitem{borghini2018listening}
G.~Borghini and V.~Hazan, ``{Listening Effort During Sentence Processing is
  Increased for Non-Native Listeners: A Pupillometry Study},'' \emph{Frontiers
  in Neuroscience}, vol.~12, p. 152, 2018,
  \url{https://doi.org/10.3389/fnins.2018.00152}.

\bibitem{zekveld2010pupil}
A.~A. Zekveld, S.~E. Kramer, and J.~M. Festen, ``{Pupil Response as an
  Indication of Effortful Listening: The Influence of Sentence
  Intelligibility},'' \emph{Ear and Hearing}, vol.~31, no.~4, pp. 480--490,
  2010, \url{https://doi.org/10.1097/AUD.0b013e3181d4f251}.

\bibitem{wang2018relations}
Y.~Wang, G.~Naylor, S.~E. Kramer, A.~A. Zekveld, D.~Wendt, B.~Ohlenforst, and
  T.~Lunner, ``{Relations Between Self-Reported Daily-Life Fatigue, Hearing
  Status, and Pupil Dilation During a Speech Perception in Noise Task},''
  \emph{Ear and Hearing}, vol.~39, no.~3, pp. 573--582, 2018,
  \url{https://doi.org/10.1097/AUD.0000000000000512}.

\bibitem{ward2017effect}
L.~Ward, B.~G. Shirley, Y.~Tang, W.~J. Davies \emph{et~al.}, ``{The Effect of
  Situation-Specific Non-Speech Acoustic Cues on the Intelligibility of Speech
  in Noise},'' in \emph{INTERSPEECH}, Stockholm, Sweden, 2017, pp. 2958--2962,
  \url{https://doi.org/10.21437/Interspeech.2017-500}.

\end{thebibliography}

\end{document}